# Using Cosmic Ray Muons to Assess Geological Characteristics in the Subsurface

H.R. Gadey[a], R. Howard[a], S.C. Tognini[b], J.L. Meszaros[b], R. Montgomery[b], S. Chatzidakis[c], J. Bae[c], R. Clark[d]

[a]Pacific Northwest National Laboratory, 902 Battelle Blvd, Richland, WA 99354 (harish.gadey@pnnl.gov)
[b]Oak Ridge National Laboratory, 5200, 1 Bethel Valley Rd, Oak Ridge, TN 37830
[c]Purdue University, 516 Northwestern Ave, West Lafayette, IN 47906
[d]Department of Energy, 1000 Independence Avenue Southwest, Washington, D.C. 20585

## INTRODUCTION

Muons are cosmogenic charged particles with an average energy of 4 GeV at sea level and are about 207 times heavier than electrons [1,2]. Muons are, therefore, naturally produced minimum ionizing particles with a range of several kilometers beneath the surface of the earth before being stopped. The subsurface range of muons is orders of magnitude greater than traditional radiation like gamma or neutrons. Subsurface characterization techniques like ground penetrating radars, borehole technologies, seismic imaging or gravity methods face several challenges with respect to the maximum achievable depth, inadequate coverage, high logistical cost, and lack of readily available expertise. Muography-based techniques might help provide a solution to some of the challenges faced by the subsurface characterization community. This detection system is planned to be deployed in the subsurface at the location of interest for benchmarking purposes. Depth-dependent muon flux measurements depend on several factors including the initial energy, zenith angle of the muon, and the density or overburden of the host rock above the detector. These variations can be used to infer properties of the subsurface such as overhead thickness or the density of the host rock without using several boreholes [3]. In addition, once this detection system is in place, the azimuth and zenith angles of the detector can be varied to capture multiple projections of the overhead. This potentially provides the ability to reconstruct the three-dimensional geometry of the facility using a single detection system fixed at a subsurface location. Since muon measurement campaigns are carried out over a long-time frame (weeks to months), potential temporal overhead density variations originating from changes in soil water content can also be monitored.

In this work, a low-power scintillation detection system is introduced. This system employs four scintillation bodies each measuring 102 cm x 51 cm x 5 cm. These scintillation bodies are fabricated to be assembled in two planes with each plane accommodating two scintillation panels. It is desired to minimize power consumption to ensure deployment at remote locations with limited infrastructure availability. The signals from the scintillation bodies are read by the Fermi National Accelerator Laboratory's (FNAL) QuarkNet data acquisition system (DAQ). This system offers a low power near-real time coincidence identification mechanism. For accurate characterization of the detection system, several simulation studies are planned to be performed. These studies will help accurately evaluate the solid angle as a function of the distance between the detector planes, improve DAQ calibration efforts, and evaluate the performance of the reconstruction algorithms.

The measured muon flux at a subsurface facility will be used to initially benchmark the new detection system and measure overhead thickness as a function of depth, zenith, or azimuth angle. This experimental data will then be used to compare against literature and published historical subsurface models. Once the system is benchmarked, this can also be deployed external to the potentially uncharacterized facility to obtain preliminary information like overhead thickness and study any bulk density changes as a function of zenith and azimuth angle. The following sections describe the detector components, readout electronics, and simulation efforts.

## DETECTION SYSTEM DESIGN

This section covers the components of the muon detection system including the scintillation panels, detector frame, and auxiliary equipment required for collecting data. The scintillation panels are responsible for detecting muons and converting the energy deposited into optical photons. These photons are converted into electrical signals using silicon photomultipliers (SiPMs). The DAQ from FNAL provides the capabilities to process the detector pulses from the SiPMs by identifying near real-time coincidence events. The detector frame is employed for providing mechanical stability to the system by maintaining a desired distance between the two detection planes and facilitating the rotation of the system along the zenith. The auxiliary equipment section describes the components and hardware required for the successful operation of the muon detection system.

### Scintillation Panels

Organic scintillators were chosen for this work to address several existing challenges with traditional inorganic scintillators, such as the crystalline structure of the material leading to brittle mechanical properties, moisture uptake over time, and possible crystal cleavage during long-distance transportation. Four custom-designed Saint Gobain scintillators are used in this work. Two detection panels are mounted in each plane and the detector geometry in both planes is orthogonal with respect to each other (Fig. 1). This ultimately leads to four detection quadrants.

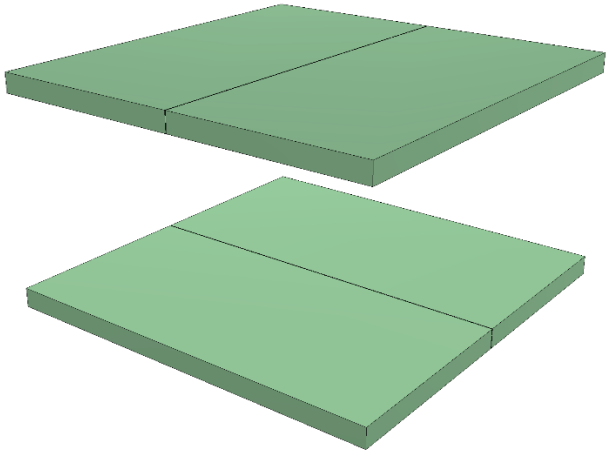

Fig. 1. Simulation rendering showing the four scintillation panels arranged orthogonally in each detection plane to realize a total of four detection quadrants.

The current design provides an active detection area of about 10,000 cm$^2$. The scintillation media is manufactured using polyvinyl toluene (PVT) which has a density of about 1.032 g/cm$^3$ and stopping power of about 1.965 MeV cm$^2$/g [4]. The expected energy deposition in each scintillator when a muon traverses it at a zero-degree zenith angle is approximately 10 MeV. Energy deposition of greater than 10 MeV can be expected when muons travel with an inclined angle (zenith angle greater than zero degrees). The optical photon yield of PVT is about 64% of anthracene (~11,100 photons/ MeV) [5]. This yields about $1.11 \times 10^5$ photons per muon interaction when it vertically enters (zero-degree zenith angle) the detection media. Since the expected energy deposition by cosmic ray muons is greater than that of background radiation, an energy threshold of 2 MeV was chosen to discriminate against background radiation. To increase the optical photon collection, multiple layers of Teflon wrapping are used. A thin aluminum layer is wrapped around the Teflon wrapping to provide electromagnetic shielding to the scintillator electronics from external sources. The exterior of the aluminum shielding is surrounded by a black vinyl layer tape to optically isolate the scintillator from the external environment.

The next step in the detection process is to convert optical photons to electrical signals. This has been traditionally accomplished using photomultiplier tubes (PMTs). In this work, we used SiPMs for several reasons including:
- low operating voltage of SiPMs compared to PMTs.
- Immunity to electric and magnetic fields.
- modular nature and compact size allowing the design of novel detection geometries.

Prefabricated OnSemi 4×4 SiPM J-Series arrays are used in this work to enable ease of integration with other electronics. The charge collected from the SiPM anodes is routed through the charge-to-voltage preamplifier that converts the current signal from the anode to a voltage signal. The SiPM array is mounted at the center of the 102 cm x 5 cm face for all scintillators (the long end). This was done to minimize the light loss from interactions taking place at the far ends of the scintillators.

A transparent silicone gel interface is utilized to optically couple the scintillation medium and the SiPM array. A black acrylonitrile butadiene styrene is used in the capacity of a carrier mounting plate. An aluminum top cap is used to protect the detector electronics from the external environment. A ground connection is established using the electrical ground screw between the electronics and the aluminum foil.

**DAQ Electronics**

The electrical signal processing and digitization were performed using the FNAL QuarkNet DAQ system as shown in Fig. 2 [6]. This system was chosen as opposed to other equipment because of its ability to identify coincidence events in near real-time in conjunction with having relatively low power consumption. It is noted that the DAQ is sensitive to only negative polarity pulses. Therefore, a design effort was undertaken at Saint Gobain to develop a custom negative polarity preamplifier solution. Using the current SiPM design, it is estimated that a 10 MeV muon interaction close to the SiPMs would yield a signal about -500mV in amplitude with a rise and fall time between 75-100 ns and 150-200 ns respectively. For a 10 MeV muon interaction, the minimum expected energy deposition at the far end of the scintillator would result in a signal amplitude reduction of about 30%. This would yield an amplitude of about -350mV. A 2 MeV muon interaction (threshold) close to the SiPM array would yield -100mV, which is still distinguishable from a -350mV amplitude resulting from a 10 MeV interaction at the far end of the scintillator.

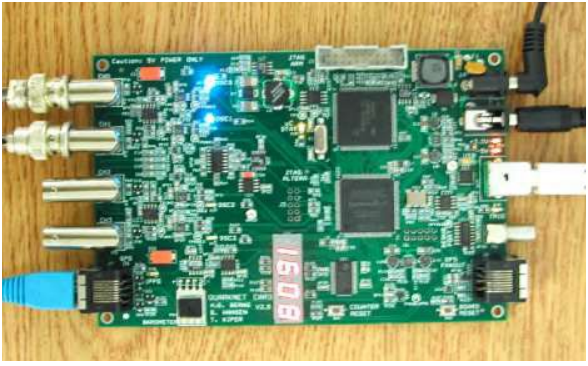

Fig. 2. Top view of the FNAL QuarkNet DAQ system.

The voltage signals from the four preamplifiers are routed through the front-end conditioning electronics on the DAQ. Discriminators are implemented in the DAQ to prevent the system from triggering on low amplitude pulses. This is initially set at -300mV by default but can be user-defined based on specific applications. The discriminator output is then fed into the time-to-digital converters (TDCs). The TDCs measure the time when the leading edge crosses the threshold and the time when the falling edge amplitude is below the threshold. In essence, this provides us with the duration of time when the signal is above the user-defined threshold. This time information is used to estimate the amplitude of the signal. Considering the same pulse shape properties, a high energy deposition will result in a signal yielding a longer time interval between the rising and falling edge above the threshold. Therefore, this information on time interval is mapped against the energy deposited by the muon in the scintillation medium. Although there are other pulse digitization techniques that are superior to the current methodology employed in terms of spectral performance, our primary objective of this work is to estimate muon flux as opposed to high-resolution spectroscopy. Therefore, the FNAL DAQ helps meet the requirements of this work. Since the minimum time interval for the time-to-digital converters is 1.25 ns, the time over threshold width is measured in multiples of 1.25 ns. Given that the rise and fall times of the Saint Gobain preamplifier are about two orders of magnitude greater than the time intervals of the DAQ, no significant pulse processing challenges are anticipated with using the FNAL DAQ. Using the muon count data, the muon flux, $\phi_\mu$, is calculated as shown in eq. 1.

$$\phi_\mu \left[\frac{\#}{cm^2 \cdot sr \cdot s}\right] = \frac{N}{A_{sc} \cdot \Omega \cdot \Delta t} \quad (1)$$

where $N$ is the total number of muons, $A_{sc}$ is the active area of the scintillator plane, $\Omega$ is the solid angle, and $\Delta t$ is the time interval of the measurement campaign.

The DAQ will be initially tested using a waveform generator to capture data and troubleshoot any challenges in the signal processing chain. This system will also be used to test the coincidence identification mechanism on the DAQ by varying the timing profile between the two pulses.

**Detector Frame and Auxiliary Components**

We chose Aluminum 6061 as our detector frame for the scintillation panels because of its favorability both in terms of the total weight and structural stability. A modular frame is being designed to facilitate ease of transportation and assembly by two or three individuals in the field. The conceptual design frame is shown in Fig. 3.

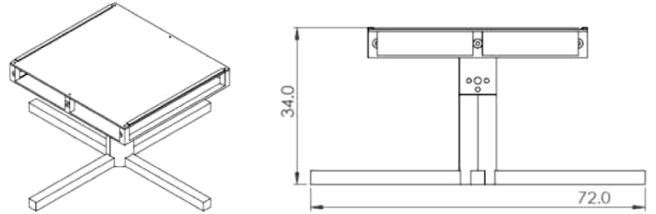

Fig. 3. Isometric view (left) and side view (right) of the conceptual muon detector frame. It is noted that all the dimensions are represented in inches.

This frame is designed to support two detection planes parallel to each other. The lip on either end of the frame is intended to hold the scintillation panels in place when it is rotated along the zenith. The planned angles of zenith rotation are in increments of 15º from 0º to 60º.

Apart from the detection frame, some of the other components that are planned to be used in this work include the Raspberry Pi device along with a touch screen user interface. This is used to replace laptops or PCs that are typically used with such systems to reduce the total power consumption of the system. A battery power system for extended operation, frame mounting, and adjusting devices are expected to be some of the additional components required to successfully operate the detection system.

**SIMULATIONS**

Simulation studies play an important role in estimating the angular resolution of the muon detection system. Generally, an increase in distance between the two planes results in improved angular resolution at the expense of increased measurement time required. On the other hand, placing two detection planes very close to each other to decrease measurement time (by increasing the solid angle) results in possible misclassification of background interactions as muons. Therefore, extensive simulation studies are being performed to determine the minimum distance between the two detection planes to achieve a given angular resolution. In this work, the Monte Carlo particle transport code Geant4 (GEometry ANd Tracking version 4) is used to characterize solid angle acceptance as a function of detector distance [7-9]. Analytically, the solid angle acceptance as a function of the face-to-face distance between the planes is presented in Table 1. It is important to note that the threshold of the detection system also plays a role in the solid angle acceptance.

TABLE 1. Analytical Estimates of Solid Angle Acceptance as a Function of Distance

| Face-to-Face Distance (cm) | Solid Angle (degrees) |
|---|---|
| 10 | 157.4 |
| 30 | 118.1 |
| 50 | 90.0 |
| 70 | 71.1 |
| 90 | 58.1 |

This simulation effort is ongoing and will help assess the solid angle resolution of the system and estimate the time required to obtain a statistically significant number of muons in field experiments. The top, side, and isometric view of the muon detection system in simulation space when a single muon interacts with two scintillation panels is shown in Fig. 4. In addition, the orthogonal orientation of the detection planes enables us to achieve preliminary geotomography capabilities.

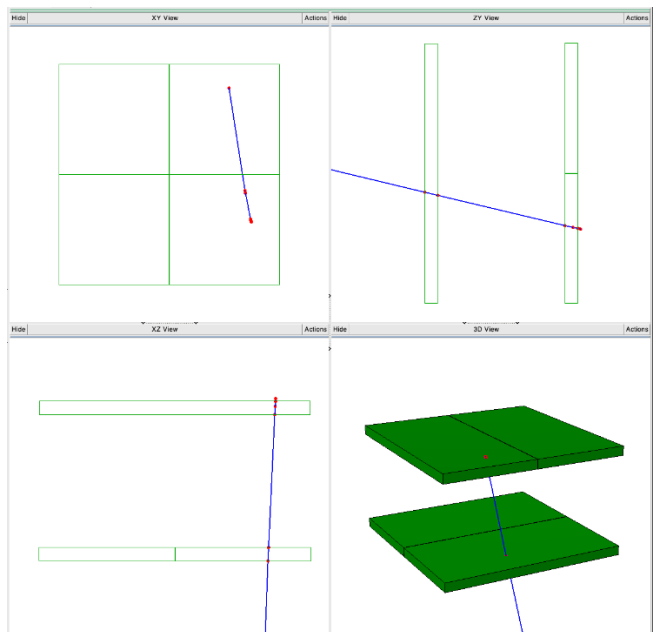

Fig. 4. Top, side, and isometric view of a single muon interaction in both scintillation detector planes.

**DETECTOR CHARACTERIZATION AND FUTURE WORK**

Once the detectors are fabricated by Saint Gobain, initial tests will be conducted at the production facility. These tests include observing the signal output to check the pulse profile and the frequency of the pulses. Light leakage tests will also be performed by directing an illumination source towards the detector and observing any unexpected increase in baseline to ensure that optimal optical isolation is achieved in the detection body.

After the detectors are delivered to PNNL, several calibration tests are planned to be performed. The first test will include testing the DAQ with all four scintillation bodies to ensure that data is appropriately collected from the detectors. This will be followed by gain matching studies that will ensure optimal signal amplitudes from all detection bodies. Then, the scintillation media will be mounted on the detector frame and calibration will be performed at various zenith angles, 0°, 15°, 30°, 45°, and 60°. Finally, the performance of a portable power source will be evaluated to provide power consumption information during operation to inform battery swap schedules.

This work will present the preliminary results from the simulation studies detailing the separation between the planes, the resulting counting time for 10,000 muons, and the anticipated angular resolution.

**ACKNOWLEDGEMENT**

This research was sponsored by the Spent Fuel and Waste Science and Technology Program of the U.S. Department of Energy (DOE) and was carried out at Pacific Northwest National Laboratory under contract DE-AC05-76RL01830 with Battelle. The authors would like to thank Kris Kuhlman, David Sassani, and Emily Stein from Sandia National Laboratories and Robert Clark and Tim Gunter from DOE for providing guidance and support to this project.